\documentclass[showpacs,floats]{revtex4}
\usepackage{amsfonts}
\usepackage{amsmath}
\usepackage{graphicx}
\usepackage{dcolumn}
\usepackage{bm}

\input{tcilatex}

\begin{document}

\preprint{}

\title{Bridging Phases at the Morphotropic Boundaries of Lead-Oxide Solid
Solutions}
\author{B. Noheda$^{1}$ and D.E. Cox$^{2}$ \\
$^{1}$ Materials Science Centre, U. of Groningen, Nijenborgh 4, 9747AG
Groningen, The Netherlands\\
$^{2}$ Physics Department, Brookhaven National Laboratory, Upton, New York
11973, USA}

\begin{abstract}
Ceramic solid solutions of PbZr$_{1-x}$Ti$_{x}$O$_{3}$ (PZT)\ with
compositions $x\simeq 0.50$ are well-known for their extraordinarily large
piezoelectric responses. The latter are highly anisotropic, and it was
recently shown that, for the rhombohedral compositions ($x\lesssim $ 0.5),
the piezoelectric coefficients were largest away from the polar direction,
contrary to common belief. Shortly afterwards a low-symmetry monoclinic
phase was observed by synchrotron x-ray diffraction at around $x=0.50$.
Similar behavior and features are also present in a number of related
lead-based strongly-piezoelectric single crystals, such as Pb(Mg$_{1/3}$Nb$%
_{2/3}$)$_{1-x}$Ti$_{x}$O$_{3}$, Pb(Zn$_{1/3}$Nb$_{2/3}$)$_{1-x}$Ti$_{x}$O$%
_{3}$, and Pb(Sn$_{1/2}$Nb$_{1/2}$)$_{1-x}$Ti$_{x}$O$_{3}$, with
piezoelectric coeficients of about 2500 pm/V, the highest values recorded to
date. Recent experimental and theoretical work has greatly improved our
understanding of these technologically-important systems, but there are
still some open questions. In this review we try to summarize the most
recent progress in the field.\bigskip
\end{abstract}
\maketitle

\section{Introduction}

\bigskip

Gen Shirane will be long remembered as one of the pioneers in the
application of neutron scattering techniques to phase transitions. It may
not be so widely appreciated that he was also a pioneer in the first studies
on the highly piezoelectric ceramics of PbZr$_{1-x}$Ti$_{x}$O$_{3}$ (PZT),
which he and his colleagues carried out in Japan in the early 1950's \cite%
{Shirane1, Shirane2, Shirane3, Sawaguchi}. When he once more took up the
study of piezoelectric systems in the last few years of his life, he was
again one of the pioneers in the renaissance of this field.

\bigskip

Over the many years that passed since his earlier work, it came to be
generally accepted that the so-called morphotropic phase boundary (MPB) in
PZT at $x\approx $ 0.50, which is associated with a number of remarkable
dielectric and piezoelectric properties, is a region of coexistence between
the tetragonal, Ti- rich (T) and rhombohedral, Zr-rich (R) phases \cite%
{Jaffe} with a variable width dependent on the homogeneity of the powders 
\cite{Isupov, Kakegawa, Mishra} and the grain size\cite{Cao}. It should be
noted that no single crystals of PZT in the morphotropic region are
available. A comprehensive review of the early research on the MPB can be
found in a recent paper by Glazer \textit{et al.} \cite{Glazer04}.

\bigskip

In the late 1990's a renewed interest in piezoelectricity was awakened due
mainly to work from the MRL at Pennsylvania State University. Park and
Shrout \cite{Park} reported on high-strain piezoelectric rhombohedral
single-crystals of Pb(Mg$_{1/3}$Nb$_{2/3}$)$_{1-x}$Ti$_{x}$O$_{3}$ (PMN-xPT)
and Pb(Zn$_{1/3}$Nb$_{2/3}$)$_{1-x}$Ti$_{x}$O$_{3}$ (PZN-xPT) with
electromechanical strain values of up to 0.6\%, little hysteresis, and
piezoelectric coefficients of about 2500 pm/V when oriented along the [001]
direction. These crystals can also show huge electromechanical strains of up
to 1.7\% under an [001] electric field related to the induced R-T
transformation \cite{Durbin}. Park and Shrout explained the lack of
hysteresis as being due to a \textit{domain engineered state} in which the
rhombohedral domains are symmetrically oriented to produce a macroscopic
tetragonal symmetry. Within each domain the polarization would rotate under
the [001] field to align closer to the field direction. At about the same
time Du \textit{et al.} showed, using a phenomenological approach, that
rhombohedral PZT also exhibits maximum piezoelectric response under an
electric field oriented along the [001] direction, and not along the polar
[111] direction \cite{Du}, which was later confirmed experimentally \cite%
{Taylor, Guo, Reszat}.

\bigskip

Shortly afterwards, a hitherto unsuspected monoclinic (M) phase was observed
at the MPB by high-resolution synchrotron x-ray powder diffraction
techniques \cite{Noheda99}. The symmetry of the new phase was found to be
very low - space group $Cm$, which has just a mirror plane and no symmetry
axis. This mirror plane is also the only symmetry element common to the
well-known R (space group $R3m$) and T (space group $P4mm$) phases, so the M phase can
be viewed as a bridge between these two structures. In the powder
diffractograms the monoclinic phase has some very distinctive features.
Although from a casual inspection of just the pseudocubic (hh0) and (hhh)
reflection profiles it might be possible to mistakenly infer the coexistence
of tetragonal and rhombohedral phases, a careful consideration of the full
pattern makes it clear that this cannot be the case. In particular, the
well-defined doublets at the pseudocubic (h00) reflections (see, \textit{e.g.%
}, Fig. 2 in Ref. \cite{Noheda00PRB}) demonstrate that the R-T coexistence
model has to be discarded. The lattice parameters and structure of the $%
x=0.48$ monoclinic phase in PZT were determined at 20 K from a Rietveld
profile analysis of the whole x-ray diffractogram \cite{Noheda00}. The full
diffraction pattern was well explained by an M ($Cm$) phase in which the
polar axis is tilted 24$^{\circ }$ away from the [001]$_{pc}$ direction
towards the [111]$_{pc}$ direction, showing the bridging role of this
intermediate phase. In subsequent work, the lattice parameters were measured
as a function of temperature and composition, and a phase diagram around the
morphotropic region was proposed, as shown in Fig. 1 \cite{Noheda00PRB}.
However, recent neutron studies have shown the presence of a few very weak
superlattice peaks which are not accounted for by this simple model, as will
be discussed in detail in the next Section.

\bigskip 
\begin{figure}[tbp]
\includegraphics[scale=0.5]{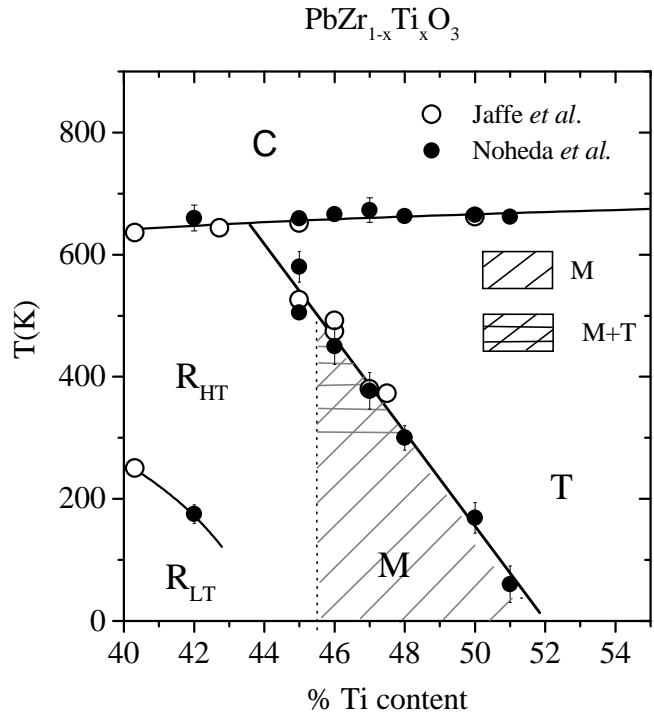}
\caption{Phase diagram of PbZr$_{1-x}$Ti$_{x}$O$_{3}$ around the MPB, as
proposed in Ref. \protect\cite{Noheda00} based on synchrotron x-ray powder
diffraction data.}
\end{figure}

\bigskip

Around the same time, Fu and Cohen \cite{Fu00}, using first-principles
calculations on BaTiO$_{3}$, reported that the high electromechanical
response observed in rhombohedral crystals under an electric field oriented
along the [001] axis was due to the rotation of the polarization within the
plane defined by [001] and [111]. This is precisely the monoclinic plane of
PZT! Thus the intermediate states induced by a field would have monoclinic
symmetry. This led to the postulation\cite{Guo,Bellaiche00} that the
presence of a monoclinic phase at the MPB was the direct cause of the great
enhancement of the piezoelectric response long known in this compositional
region.

\bigskip

First-principles calculations by Bellaiche \textit{et al.} \cite{Bellaiche00}
reproduced the stability of the monoclinic phase in a narrow region at the
MPB when a random Zr/Ti cation distribution was taken into account. The
stability of a monoclinic phase of space group $Cm$ (the so-called M$_{A}$
phase) in between the R and T phases was also explained using a
Landau-Devonshire approach by Vanderbilt and Cohen\cite{Vanderbilt}, who
furthermore predicted two other monoclinic phases (M$_{B}$ and M$_{C}$)
bridging the rhombohedral and orthorhombic, and the orthorhombic and
tetragonal phase of perovskites, respectively. These two latter monoclinic
phases have since been observed in PMN-xPT and PZN-xPT\cite{Singh01,
Xu01,Cox01,Ye01,Yamada02,Kiat02, NohedaPMNPT, Ora02, Uesu02, Dkhil02,
Singh03, Zekria,Renault05} and, more recently in Pb(Sc$_{1/2}$Nb$_{1/2}$O$%
_{3}$)$_{1-x}$Ti$_{x}$O$_{3}$ (PSN-PT)\cite{Haumont03, Haumont05}.

\bigskip

An earlier review can be found in Ref. \cite{Noheda02}, but since
then numerous papers have been published on the topic and new ideas have
been advanced. Due to the intrinsic cation disorder in these lead oxide
solid solutions, the average structure differs from that at the level of the
unit cell. Small domains and strain effects can result in a markedly
different picture also at the mesoscopic scale. In general, one can say that
current work on the topic is directed towards the observation of the MPB
regions by different techniques and thereby obtain a consistent picture of
the phenomena observed at the different length scales. Furthermore, there
exist also concerns about the stability of the intermediate monoclinic phases%
\cite{Amin04,Kisi}, as well as about whether the high electromechanical
response of these compounds is mainly intrinsic or extrinsic. These issues
are still not fully resolved but there has been substantial progress in the
field. In the following sections we will try to give a general idea of the
most recent developments.

\section{\protect\bigskip Structural Instabilities at the MPB}

\subsection{Oxygen Octahedral Tilts}

First-principles calculations by Fornari and Singh \cite{Fornari} have shown
that, in ferroelectric perovskites, strain can induce local
anti-ferrodistortive tilts of the oxygen octahedra in coexistence with the
ferroelectric cation displacements. Such tilts would be expected to lower
the symmetry of the lattice, and indeed evidence of a cell-doubling
transition in PZT with $x=0.48$ at low temperatures was reported by Ragini 
\textit{et al.} \cite{Ragini01} from electron diffraction measurements, and
by Noheda \textit{et al.} \cite{Noheda00PRB} from neutron powder diffraction
data even though no indication of such a transition was observed in the
corresponding x-ray powder patterns due to the low scattering power of
oxygen.

\bigskip

However, detailed Rietveld analysis of the neutron powder data from the $%
x=0.48$ composition has proved to be far from straightforward. Although the
reported raw data appear to be very similar at first sight, there are currently three
different interpretations. The neutron patterns are characterized by the
appearance of one or more weak superlattice peaks, indicating the doubling
of the monoclinic unit cell along the $c$-axis, as reported in Ref. \cite%
{Noheda00PRB}, and by Ranjan and coworkers \cite{Ranjan, Hatch}, who
concluded the low-temperature structure was also monoclinic, space group $Cc$%
, with a doubled cell characteristic of antiphase tilting of the oxygen
octahedra. On the other hand, Frantti \textit{et al.} \cite{Frantti02} have
concluded that the weak superlattice peaks should be attributed to a
minority rhombohedral phase with $R3c$ symmetry in coexistence with the
monoclinic $Cm$ phase. However, this conclusion was not supported by
electron diffraction data obtained by Noheda \textit{et al.} \cite{NohedaTEM}%
, which showed no evidence of a rhombohedral phase, but instead the $Cm$
phase in coexistence with a minority $Cc$ phase. Furthermore, a similar
coexistence model was deduced from a recent Rietveld analysis of the neutron
data mentioned in Ref. \cite{Noheda00PRB}, namely monoclinic $Cm$ and $Cc$
phases in the approximate ratio 4:1 \cite{Cox05}. At about the same time,
Ranjan \textit{et al.} \cite{Ranjan05} reported that the previously
published neutron data \cite{Ranjan, Hatch} also favored a coexistence
model, but with a markedly different ratio of $Cm$ and $Cc$ phases, in this
case about 1:2. In both studies it was concluded that a minority $R3c$ phase
fails to account for the superlattice peak positions, but this is disputed
by Frantti \textit{et al.} \cite{Frantti05}, who argue that if anisotropic
peak broadening is allowed for correctly \cite{Frantti03}, a coexistence
model of $Cm$ and $R3c$ phases accounts well for the peak positions and
intensities.\cite{Footnote} In a very recent paper, published during the
writing of this manuscript, Woodward \textit{et al.}\cite{Woodward} have
presented a detailed analysis of TEM measurements across the phase diagram
of PZT and propose a single Cc phase around the MPB at sufficiently low
temperatures. These authors point out that the monoclinic Cc phase supports
the tilt system $a^{-}b^{-}b^{-}$ (in Glazer's notation\cite{Glazer72}),
which is more general than those previously considered. If so, the Cc phase
can then be considered as the structural bridge between the R3c ($%
a^{-}a^{-}a^{-}$ tilted ) and the untilted Cm phase.\cite{Woodward}

\bigskip

Thus it must be said that the current situation is rather confusing, and
illustrates the difficulty of obtaining a definitive result from Rietveld
analysis alone in these complicated piezoelectric systems, which are
typically characterized by low symmetry, local order, pronounced
microstructural effects, and coexistence of two (or more) closely-related
phases. Nevertheless, we believe that although oxygen tilts may play an
important role in lowering the elastic energy of the system, and in
stabilizing the monoclinic phase\cite{Fornari}, it is the symmetry element
of the cation displacements (\textit{i.e.} the mirror plane of the simple
monoclinic phase, $Cm$) which defines the symmetry of the polarization and
the anisotropy of the physical properties.

Further insight into the structural relationships in the PZT system has been
obtained from diffraction experiments carried out under hydrostatic
pressure. From synchrotron x-ray powder data and Raman spectroscopy studies,
aided by first-principles calculations, of a sample with $x=0.48$ at room
temperature, Sani \textit{et al.}\cite{Sani} have shown that polarization
rotation from the T towards the R phase, through the monoclinic M$_{A}$ phase, 
occurs under pressure, together with tilting of the oxygen octahedra in agreement with Ref. \cite{Fornari}.
Neutron and x-ray measurements on a series of PZT compositions by Rouquette
and colleagues \cite{Rouquette04} have confirmed both the rotation within
the M$_{A}$ phase and the oxygen tilts; in particular, for $x=0.60$ there is
a transition from T to M$_{A}$ symmetry around 1 GPa, followed by a
transition to the antiferrodistortive $Cc$ structure at about 3 GPa. In a
subsequent neutron study of an $x=0.48$ sample \cite{Rouquette05}, a series
of monoclinic ($Cm$ and $Cc$) and triclinic ($F1$ and $F\overline{1}$)
phases were reported, and the data suggest that a rhombohedral phase is
never reached. Recently Wu and Cohen have reported that a T-M-R-C sequence
can be induced by pressure in pure PbTiO$_{3}$. Thus hydrostatic and
chemical (Zr-substitution) pressure have a similar effect on PbTiO$_{3}$,
and a MPB in between the T and R phases is predicted at about 12 GPa at low
temperatures \cite{Wu&Cohen}.

\subsection{Electric Field Effects on the MPB}

To investigate the effect of an electric field on the behavior of the MPB,
x-ray diffraction patterns from ceramics of rhombohedral and tetragonal PZT
were measured before, during and after poling. The effect of the poling was
found to be irreversible and consistent with the induction of an M$_{A}$
type distortion\cite{Guo}. No elongation of the unit cell along the polar
directions was observed. The piezoelectric coefficient measured from the
patterns of the rhombohedral composition along [001] was 500 pm/V,
surprisingly, about equal to that obtained by dilatometry. The availability
of single crystals of PMN-xPT\ and PZN-xPT, has allowed the induced symmetry
changes to be followed directly \textit{in-situ} under an E-field, and
showed that close to the MPB the initial R phase was indeed irreversibly
transformed into M (M$_{A}$ or M$_{C}$) by applying an [001] E-field, and
that the monoclinic phases acted as bridging phases in the
rhombohedal-tetragonal transformation not only as a function of composition
but also under an applied electric field \cite{Noheda01PRL, Noheda02PRB,
Uesu02, CaoJAP05}, in excellent agreement with theoretical calculations\cite%
{Bellaiche01}. Viehland and co-workers showed that this behaviour was not
specific to the [001] oriented crystal, and that a [101] E-field could
induce the rhombohedral-orthorhombic transformation and giant piezoelectric
deformations as well\cite{Viehland01, CaoPRB05}. In particular, Cao \textit{%
et al.}\cite{CaoPRB05} have shown that electric field oriented in this
direction can induce the rarely-observed monoclinic M$_{B}$ phase, predicted
by Vanderbilt and Cohen\cite{Vanderbilt}

\subsection{Stability Issues at the MPB}

\bigskip As mentioned above, theoretical calculations fully support the
stability of the M phase near the MPB under zero field and further away from
the MPB under a properly oriented electric field. Renault \textit{et al.}
have checked the field-induced monoclinic phase in PZN-9PT monodomain
crystals after removal of the field, finding that the monoclinic phase is
stable and the samples do not show partial depoling \cite{Dammak02}. Topolov
has interpreted the primary role played by the monoclinic phase as elastic
matching and strain relief during the R-T transformation \cite{Topolov01,
Topolov04}, which has found direct experimental support from Frantti \textit{%
et al.}\cite{Frantti04}. However, there is in the experimental literature
some divergence concerning the stability of the M phase. Until now, an
indication of the monoclinic phase in PZT has been reported by several
techniques\cite{Lima, Guarany, Eichel}, but it is fair to say that the
evidence is somewhat elusive. Only high-resolution synchrotron or neutron
diffraction measurements at low temperatures on high quality powders are
able to show the M phase definitively due to the small distortion with
respect to the tetragonal unit cell. Coexistence with the R and T phases has
often been reported, but it is primarily in very homogeneous samples with
compositions around $x=0.48$ at low temperatures where the M phases ($Cm/Cc$%
) show up most distinctly. X-ray diffraction from laboratory sources has
most often not been capable of directly resolving the subtle M splittings,
and the best evidence has been provided by the improved fits obtained from
Rietveld analysis of the whole profile when the M phase is considered\cite%
{Ragini01, Ragini02}. Moreover, the stress dependence of the M phase
discussed above explains why in certain ceramics with relatively small grain
size and relatively large microstrain, the low temperature monoclinic phase
was not present and, instead, a rhombohedral phase was observed\cite%
{NohedaFerro}.

\bigskip

As will be discussed later, it is worth noting that, due to the
near-degeneracy of the different phases at the MPB, small differences in the
strain state of the samples can easily alter the phase stability and thus
explain the reported divergence in the experimental results. Accordingly,
special attention must be given to sample preparation. In principle, the
structure of PMN-xPT or PZN-xPT single crystals can be determined by
crushing a small piece of the crystal to a fine powder, but to obtain a
suitably orientationally-averaged powder sample is likely to result in peak
broadening because of the induced strain. Although the M phases might still
be present\cite{NohedaFerro02}, such broadening can make it very hard to
resolve the small monoclinic splittings in the diffraction pattern. One
compromise which was found to yield extremely narrow and well-resolved
peaks, and hence to allow an accurate unit cell determination involves
careful crushing and sieving of an $\sim $40 $\mu m$ fraction\cite{Cox01,
NohedaFerro02}. However, there is the risk with this technique that the
powder averaging may not be sufficiently good to permit a reliable analysis
of the atomic positions. Thus it now appears to be widely accepted that,
given the delicate energy balance at the MPB and the large microstrains
involved, the microstructure of the materials will be a decisive factor in
the phase stability of the different phases around the MPB, explaining the
relatively diverse results in the literature, including the dependence with
grain size of the tetragonal-monoclinic ratio \cite{Bertram03}, and the
monoclinic-rhombohedral transformation induced by grinding\cite%
{NohedaFerro02}. Moreover, it has been shown that in single crystals of
PZN-PT and PMN-PT the structure of the outermost region of the crystal is
different from that of the inner part\cite{Xu103,Xu203,Gehring04,Colon04}.
This `skin effect'is responsible for some striking differences observed
between neutron and x-ray (at both medium and low energies) diffraction
data due to the very different penetration depths\cite%
{Noheda01PRL,Ohwada,NohedaFerro02}, and could also be the cause of the
differences encountered between single-crystals and ceramics. An extensive
review of the skin effect in piezoelectrics crystals is presented by Xu 
\textit{et al.} in this issue.

\section{\protect\bigskip Local versus average symmetry lowering}

Theoretical investigations point to the importance of cation disorder in
explaining the features of the MPB\cite{Bellaiche00, BellFerro05}. Because
of the large coherence length of neutrons and x-rays, diffraction data from
these probes are only capable of determining average distortions. However,
because information about static and dynamic disorder in the average
structure, or short-range order (shorter than the coherence length of the
probe), is contained in the so-called anisotropic displacement factors, or
thermal parameters, useful information can be gained about the local
structure by carefully looking at these parameters\cite{Malibert}. That the
rhombohedral phase of PZT involved more than just a long-range rhombohedral
distortion was first pointed out by Glazer \textit{et al. }\cite{Glazer78},
when they observed that the anisotropic displacement factors for the Pb
atoms determined from the neutron diffraction patterns of PbZr$_{0.9}$Ti$%
_{0.1}$O$_{3}$ were unphysically large and highly anisotropic. They also
found, in agreement with earlier work by Sawaguchi\cite{Sawaguchi}, that
they had the shape of ellipsoids flattened perpendicularly to the polar
[111] direction. This disorder seemed to be static as it showed little
change with temperature. Similar local shifts of Pb atoms were also found by
Teslic \textit{et al.} using a pair-distribution function analysis \cite%
{Teslic}. Following up on these findings, in 1998 Corker \textit{et al.}\cite%
{Corker98} proposed a model in which the Pb atoms randomly occupy three
different sites shifted, respectively, along pseudocubic [100], [010] and
[001] with respect to a common [111] displacement. The random occupancy of
these three sites leaves the average shift along the [111] direction
unchanged. Thus the average long-range rhombohedral symmetry is maintained,
but at the level of the unit cell the symmetry is lowered, as illustrated in
Fig. 2.

\bigskip 
\begin{figure}[tbp]
\includegraphics{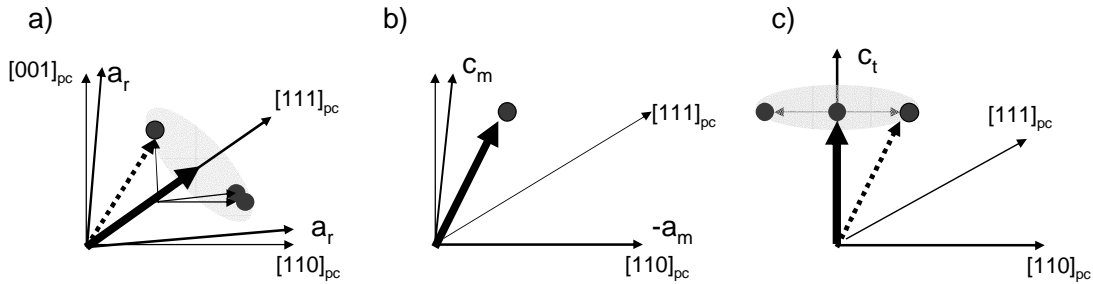}
\caption{Monoclinic plane of PbZr$_{1-x}$Ti$_{x}$O$_{3}$, containing the
rhombohedral and tetragonal polar axes. Pb shifts from the high-symmetry
positions are sketched as inferred from Rietveld analysis of synchrotron and
neutron diffraction powder diffraction data, showing: (a) The three
equivalent sites proposed by Corker \textit{et al.} \protect\cite{Corker98}
to explain the disorder in the rhombohedral phase (b) Pb shifts in the
monoclinic phase (c) four equivalent sites in the tetragonal phase. Figure
adapted from \protect\cite{Noheda00}.}
\end{figure}

\bigskip When the monoclinic phase was observed shortly afterwards in PbZr$%
_{0.52}$Ti$_{0.48}$O$_{3}$ at low temperatures\cite{Noheda99,
Noheda00}, it was found that the monoclinic displacements corresponded very
closely to one of the three sites, \textit{i.e.} [111] + [001], proposed by
Corker \textit{et al.} to explain the local disorder in the rhombohedral
phase \cite{Corker98}. The extra [001] shift with respect to an average
rhombohedral phase can equally well be viewed as a distortion from
tetragonal symmetry by rotation of the polarization away from [001] towards
[111], so that the monoclinic phase provides the structural bridge that
allows an average rhombohedral structure to change continuously to an
average tetragonal structure through the MPB. The values obtained for the
anisotropic displacement factors in the M phase did not indicate any
substantial differences between local and average order in the monoclinic
composition. This was not the case, though, for the tetragonal phase
observed at 325 K in this sample, for which strongly anisotropic
displacement factors were observed perpendicular to the polar [001] direction%
\cite{Noheda00}.

\bigskip

If the model of Corker \textit{et al. }is adapted to the tetragonal phase,
it is necessary to consider four equivalent sites that, randomly occupied,
would allow the average tetragonal structure to be retained. When these
sites were chosen to lie along [110], [1$\overline{1}$0], [$\overline{1}$10]
and [$\overline{1}\overline{1}$0] the Rietveld refinement was significantly
improved, and reasonable displacement factors were obtained, analogous to
the improvement found for the rhombohedral phase \cite{Corker98}. Any one of
these four shifts by itself would result in a monoclinic distortion, and
thus provide a structural bridge linking the tetragonal to the rhombohedral
structure by polarization rotation. This led to the proposal \cite{Noheda00}
that, as the Ti content is changed across the MPB, one of the local sites of
the R or the T phase is preferred, resulting in a monoclinic phase in a
narrow intermediate region of composition.

\bigskip

Neutron studies by Dmowski \textit{et al.} using the atomic
pair distribution function technique (PDF), which provides details of the
distribution of atomic density as a function of interatomic distance without
symmetry assumptions, have shown that the local changes in PZT across the
MPB are very gradual\cite{Dmoswki}. In particular, from a comparison of the
PDF of PZT\ with that of pure PbTiO$_{3}$ and pure PbZrO$_{3}$, it was
suggested that the local environment of the Zr/Ti cations remains relatively
invariant of composition. They found that Ti is always ferroelectrically
active across the phase diagram, while Zr is not. The variation with
composition is thus due mainly to the Pb cations, which are locally
displaced along the \TEXTsymbol{<}100\TEXTsymbol{>} and \TEXTsymbol{<}110%
\TEXTsymbol{>} directions, as in PbTiO$_{3}$ and PbZrO$_{3}$, respectively,
in order to form four, instead of three, short Pb-O bonds (as would result
from \TEXTsymbol{<}111\TEXTsymbol{>} displacements). The most significant
change with Zr-Ti content was the distribution of those displacements, 
\textit{i.e.}, from a majority of \TEXTsymbol{<}110\TEXTsymbol{>}
displacements close to $x=0$ to a majority of \TEXTsymbol{<}100\TEXTsymbol{>}
displacements close to $x=1$. There are similarities between the model of
Dmowski \textit{et al. }for the MPB and that discussed earlier, and both
propose gradual local changes across the MPB. The main difference between
the two is that while the anisotropic displacement factors show an increase
in order at the MPB, the PDF studies seem to indicate a larger amount of
disorder at the MPB.

\bigskip

By means of x-ray and neutron diffraction measurements on the related solid
solution PSN-PT, Haumont \textit{et al.} have also observed that the R-T\
transformation as a function of Ti content is gradual\cite{Haumont03}. These
authors have proposed that for low titanium contents, there is a macroscopic
rhombohedral state with local monoclinic symmetry resulting from the
combination of Pb and Sc/Nb/Ti shifts along the [001] and [111] directions,
respectively. Since, as in the synchrotron studies of PZT, the disorder is
observed to decrease from the rhombohedral side towards the MPB, it is
suggested that, with increasing Ti content, competition between cations
leads to an increase in the coherence length of the short-range monoclinic
phase, which becomes long- range in the morphotropic region.

\bigskip

Recently, Glazer \textit{et al.} have carried out detailed electron diffraction
measurements on PZT across the MPB\cite{Glazer04}. The diffuse scattering in
these patterns clearly shows the change from short-range order to long-range
order as the MPB is approached from both the Zr-rich and the Ti-rich sides.
The same authors identify the nature of this short-range order as due to
ordered cation displacements rather than compositional inhomogeneity,
because of the absence of diffuse scattering along high-symmetry directions.
Based on Rietveld profile analysis of x-ray powder data, Ragini \textit{et
al.}\cite{Ragini02} have gone further in suggesting that the nominally
rhombohedral compositions with $0.47\geq x\geq 0.38$ are more likely to be
`truly' (long-range as well as short-range ordered) monoclinic. As Glazer 
\textit{et al.}\cite{Glazer04} have pointed out, a model of
uncorrelated monoclinic unit cell distortions throughout the phase diagram
with correlation lengths increasing towards the MPB, provides a more
plausible explanation of the rather diverse experimental findings concerning
the details of the phase stability and coexistence, since these would be
very much dependent on the sample homogeneity, and even more on the
coherence length of the experimental probes. Overall, this picture can
explain the lack of a well-defined R-M phase boundary (indicated by the
dotted line in Fig. 1), or any piezoelectric and dielectric peaks for
that composition (x $\simeq $ 0.455).

\bigskip

This model is also attractive in considering the effect of an electric field
in these systems. When an [001]-oriented field is applied to the so-called
rhombohedral phase, the [001] local shift will be stabilized with respect to
the other two, and a long-range monoclinic phase will be induced through
polarization rotation\cite{Guo}. In this case the macroscopic field-induced
rhombohedral-to-monoclinic phase transition would actually correspond to a
short-range-to-long-range monoclinic phase transformation, and the electric
field would have the effect of increasing the correlation length. Guo 
\textit{et al.}\cite{Guo} showed that the piezoelectric elongation
of rhombohedral PZT\ along [111] was undetectable and also that the
piezoelectric elongation of tetragonal-monoclinic PZT ($x=0.48$ at room
temperature) was minimal along the [001] direction. These results are
consistent with the stabilization of one of the local monoclinic
displacements by a properly oriented field on both sides of the MPB.

\bigskip

Using density-functional theory calculations, Grinberg \textit{et al.}\cite%
{Grinberg02} have shown that in PZT (and similarly in PMN-xPT\ and PZN-xPT%
\cite{Grinberg04}) Pb \TEXTsymbol{<}111\TEXTsymbol{>} displacements are
indeed avoided, as indicated by the experiments. The effectiveness of the
method to deal with related Pb and Bi complex oxides is illustrated by its
predictive power\cite{Grinberg05}. The calculations also reveal that the
Pb-shifts deviate from the average polar direction, shifting towards the
neighbouring Ti atoms and away from the Zr atoms, due to the increased
repulsion with the larger cations (see Figure 3), while the shifts of the
Zr/Ti atoms do not change significantly with composition. A majority of
Ti-rich, Zr-rich or neutral faces in the material, towards which the Pb can
shift, will therefore decide whether the average symmetry is tetragonal,
rhombohedral or monoclinic, respectively. According to these results,
maximum disorder exists in the rhombohedral phase, with a substantial
decrease in the monoclinic phase, as experimentally observed. But, in
contrast to Refs. \cite{Glazer04} and \cite{Noheda00}, the calculations show
minimum disorder in the tetragonal phase.

\bigskip
\begin{figure}[tbp]
\includegraphics{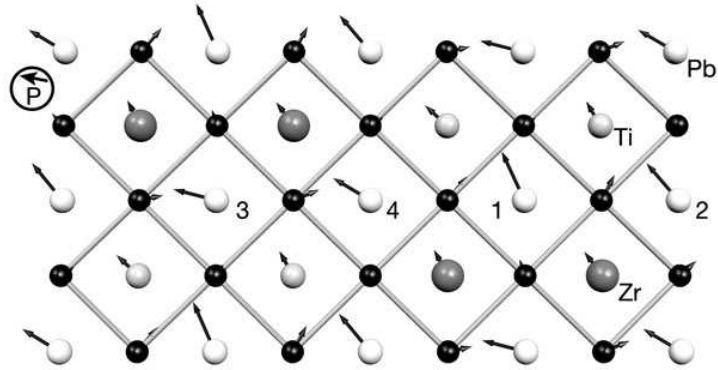}
\caption{Local displacements in PbZr$_{1-x}$Ti$_{x}$O$_{3}$ resulting from
Density Functional Theory calculations free from symmetry constraints by
Grinberg \textit{et al.}\protect\cite{Grinberg02}(reproduced with Nature's
and authors' copyright permission)}
\end{figure}
\bigskip

The latter result is in agreement with the work by Haumont \textit{et al.}%
\cite{Haumont03} on PSN-PT and that of Ragini \textit{et al.}\cite{Ragini02}
on PZT, who have carried out Rietveld analysis of neutron and x-ray
diffraction patterns of several compositions across the MPB. In contrast to
the previous synchrotron work described above performed on tetragonal PZT
with $x=0.48$ at 325 K \cite{Noheda00}, at room temperature Haumont \textit{%
et al.} and Ragini \textit{et al.} did not find highly flattened
displacement factors in the tetragonal compositions, indicative of little
difference between the local and average order on the tetragonal side of the
MPB. Molecular-dynamics simulations performed by Shin and collaborators in
equilibrated 4x4x4 supercells of PbTiO$_{3}$ at 300 K, give relatively large
standard deviation values for Pb displacements (0.105\AA , 0.102\AA , 0.099%
\AA ), corresponding to slight flattening perpendicular to the polar axis,
indicating that the origin of the local displacements observed at 325 K in
morphotropic PZT could be partially dynamic\cite{Rappe}. A true phase
boundary between the M and T phases, where a second-order phase transition
is predicted\cite{Vanderbilt}, would define the MPB as a critical line, and
the only quasi-vertical boundary, in the $x-T$ diagram. This possibility is
very attractive, since both elastic softening and the largest anomalies in
the piezoelectric and dielectric responses are expected at a second-order
phase transition, while still allowing continuous rotation of the
polarization.

\bigskip

Continuous rotation of the macroscopic polarization across the MPB due to
the rotation of the Ti-shifts in PZT has recently been proposed by Cao 
\textit{et al.} \cite{Cao2} based on x-ray absorption fine structure (XAFS)\
data. These data indicate a continuous change in the Ti local displacements
from [001] to [111] \ for compositions between $x=0.55$ and $x=0.40$. In
agreement with Corker \textit{et al.} \cite{Corker98} and Dmowski \textit{et
al.} \cite{Dmoswki}, these authors find no evidence of substantial
ferroelectric shifts of the Zr atoms. These results thus support the idea
that the local Ti dipoles follow the average structural distortion and that
the direction of the \textquotedblleft local\textquotedblright dipole
moments and the macroscopic polarization pretty much coincide.

\bigskip

Another very appealing interpretation of the monoclinic phase has been given
by Viehland, Jin and coworkers \cite{Viehland00, Jin03}. Their model is
based upon the assumption that, for such a polarization rotation mechanism,
associated with a very small domain wall energy\cite{Park}, since the
different polarization orientations are nearly degenerate, the ferroelectric
phase transforms into an inhomogeneous microdomain state which is
macroscopically homogeneous, a so-called \textit{adaptive phase}. This state
is described as a miniaturized tetragonal microdomain structure governed by
lattice accommodation under stress and electric field. Such an adaptive
phase, formed by plates containing twin-related domains, would be observed
as a homogeneous monoclinic phase by diffraction measurements. In order to
accommodate the stress and avoid misfits along the domain boundaries,
certain relationships between the lattice parameters of the tetragonal phase
($a_{t}$, $c_{t}$)\ and the adaptive phase ($a_{ad}$, $b_{ad}$, $c_{ad}$)
need to be fulfilled; in particular: \ $a_{ad}+c_{ad}=a_{t}+c_{t}$ and $%
b_{ad}=a_{t}.$ Moreover, in the M$_{C}$ phase, $a_{ad}$ and $c_{ad}$ would
form an angle $\beta =90^{o}+2A\omega (1-\omega )\phi $, where $2\phi $\ is
the angle formed by the two twin variants in the tetragonal phase, with
polarizations along [100] and [001], respectively, $\omega $ and (1-$\omega )
$ are the volume fractions of the two variants, and A is a fitting parameter
related to the volume fraction of domain walls or the size of the domains 
\cite{Wang05}. If neither an electric field nor uniaxial stress is applied,
then the stress accommodation relations between the tetragonal and the
parent cubic phase ($a_{c}$) would also give: $\ \ a_{ad}=a_{c}$. Adaptive
phases of this sort have been previously observed in martensites \cite%
{Shapiro89}. The remarkable agreement between the above relationships and
the experimental data reported for the monoclinic M$_{C}$ phases in PMN-xPT
and PZN-xPT is illustrated in Refs. \cite{Jin03,Wang05}. According to this
model the effect of an applied electric field would be to detwin the crystal, 
which would modify the relative fraction of domains
with polarizations along [001] and [100] (changing $\omega $), effectively producing the rotation
of the macroscopic polarization\cite{Wang05}. The adaptive model can explain
characteristics of the monoclinic M$_{C}$ phase previously considered as
curiosities, such as the $b_{m}$ lattice parameter being a continuous
extrapolation of the $a_{t}$ lattice parameter. Furthermore, these authors
state that a similar approach to rhombohedral domains yields the monoclinic M%
$_{A}$ phase observed in PZT\ and PZN-4.5PT. Experimental evidence of a high
domain-wall density and domain structures at various length scales at the
MPB of PMN-xPT observed by piezo-force microscopy has recently been reported
by Bai \textit{et al.} \cite{BaiAPL05}. Moreover the nanotwin model of the
monoclinic phase is also consistent with the large anisotropic broadening of
the diffraction profiles around the MPB (with narrow 0k0 peaks) observed in
PMN-xPT\cite{NohedaPMNPT}, as well as the phase coexistence often
encountered over large temperature and compositional ranges\cite{Wang05}.

\bigskip

Topolov and Ye \cite{Topolov04, TopolovYe04} consider a model of
interpenetrating phases to explain the MPB\ of PMN-xPT. The phases are
separated by planes with zero net strain. According to this model each
sample consists of two phases and two different types of regions, each one
containing both phases. In one type of region phase-1 inclusions exist in a
phase-2 matrix, and \textit{vice versa} for the second type of region. The
elastic matching at the phase boundaries relates the lattice parameters of
the intermediate monoclinic phase with those of the tetragonal and
rhombohedral phases. The same elastic matching can be obtained in different
ways; for example, with the coexistence of three phases in a more narrow
composition region. This model could account for the two different phase
diagrams reported for PMN-xPT \cite{NohedaPMNPT, Singh03, Zekria}. In common
with Viehland \textit{et al.}, this model explains the stability of the M
phase at the MPB by stress accommodation at the interfaces of the R and T
phases, as well as the relation between the lattice parameters of the
monoclinic phase and those of the neighbouring phases. But, contrary to
Viehland \textit{et al.}, it does not assume subnanometer domains of those
phases but instead a true M$_{C}$ phase in coexistence with at least one
other phase. It is evident that more experimental work is needed in order to
clarify these various pictures, and to find whether a single model applies
both to PZT and to the relaxor single crystals.

\section{\protect\bigskip Piezoelectric response}

In recent years much effort has been devoted towards the control of the
performance of these technologically-important materials \cite%
{Park,Viehland,Trolier,TrolierMuralt,Li05}, but despite all the recent
progress in understanding their fundamental behavior, the origin of the
piezoelectric response in these materials is still a matter of debate and
there is still some controversy about whether the piezoelectric properties
of these materials are intrinsic or extrinsic in nature, if they are due to
the ferroelectric domains or to the motion of the domain walls.

\bigskip

The adaptive theory described in the previous section suggests that the
large electromechanical response in these materials around the MPB is due to
detwinning, that is movement of domain walls\cite{Wang05}. A [001]-oriented
electric field would increase the size of the tetragonal twins with
polarization along the [001] direction, reducing those along the [100]
direction, until they become visible to diffraction probes. During this
process the diffraction probes will detect the rotation of the polarization
within a M phase and a M-to-T transformation. However, Damjanovic and
coworkers\cite{Damjanovic03} observed that, in PMN-33PT, the piezoelectric
response is mainly intrinsic due to shear piezoelectricity, or polarization
rotation, inside the domains, and that the multidomain state had a
relatively small effect (\TEXTsymbol{<} 20 \%) in the total response. This
is again contrary to the analysis by Renault \textit{et al.} for PZN-9PT 
\cite{Renault}, who propose that a large density of domain walls are present
in these crystals, resulting in an important extrinsic component to the
piezoelectric response. In agreement with the latter proposal are the
experiements of Wada \textit{et al.}\cite{Wada} showing that the
piezoelectric response of KNbO$_{3}$ can be considerably improved by
reducing the domain size. On the other hand, polarization switching observed
in PZN-4.5PT by Davis \textit{et al.} suggests that there is no domain
growth but continuous renucleation instead \cite{Davis05}. For small
opposing fields a quick polarization response occurs which is both
reversible and non-dissipative (indicating no activation barrier),
consistent with polarization rotation. At larger fields, nucleation rather
than domain growth is observed. The broad distribution of times suggests
nucleation across the crystal, probably at defect sites \cite{Julian}.

\bigskip

As discussed above, first-principles calculations point to polarization
rotation of R crystals under an [001]-electric field between the R and T
phase following several possible paths \cite{Fu00,Bellaiche01}. These
calculations predict intermediate monoclinic phases, and even a triclinic
phase, to be stabilized by the electric field during the rotation. These
tetragonal, monoclinic and rhombohedral phases at the MPB are nearly
degenerate\cite{Bellaiche00}, corresponding to flat energy surfaces, which
explains why subtle differences in sample preparation or microstructure can
lead to changes in phase stability. Phenomenological calculations are
consistent with this picture\cite{Vanderbilt}. In the light of these
calculations Vanderbilt and Cohen suggested that the presence of monoclinic
phases in ferroelectric perovskites arises as a consequence of unusually
anharmonic free energies \cite{Vanderbilt}. Highly anharmonic potentials
have indeed been reported for lead perovskites\cite{Kiat00}. The soft
potentials at the MPB are responsible for the non-collateral anomalies in
the dielectric and piezoelectric responses observed at the MPB, the
so-called transverse instabilities\cite%
{Bellaiche00,Ishibashi,Iwata,Wu,Bungaro,Budimir03,DamjanovicAPL02,
BudimirAPL04,Budimir05}. The presence of a ferro-ferro phase transition,
such as that occurring at the MPB's or the tetragonal-orthorhombic
transition in BaTiO$_{3}$ with a change of polarization direction, is
accompanied by a dielectric softening in the perpendicular direction\cite%
{Du,Viehland01}, associated with a shear component of the piezoelectric
tensor. An extensive review by D. Damjanovic on this topic has been
published during the writing of this manuscript\cite{Damjanovicreview05}.

\bigskip

From first-principles calculations on a tetragonal PZT crystal with random
Zr/Ti distribution, Bellaiche \textit{et al.} \cite{Bellaiche00} have shown
that in the proximity of the T-M boundary the shear piezoelectric
coefficient $d_{15}$, which is related to the rotation of the polarization
from [001] to [111], is greatly enhanced, thus explaining the observed
piezoelectric response. More recently, Wu and Krakauer were able to
calculate the electromechanical response in an ordered monoclinic phase and
obtain very large piezoelectric coefficients, similar to the experimental
ones\cite{Wu}. This explicitly shows that polarization rotation alone can
explain the observed response at the MPB. As discussed by these authors,
although this mechanism could be generalized to other ferroelectrics such as
BaTiO$_{3}$ (keeping in mind that, as pointed out in Ref. \cite{Kisi}, the
symmetry of a rhombohedral crystal under a [001] field, or a tetragonal
crystal under a [111]-field, is monoclinic in general), it is only at the
MPB where the energies of the different phases are close enough, or the
energy surfaces flat enough, for the symmetry of the crystal to change under
the available E-field. Hydrostatic pressure, as discussed above\cite%
{Wu&Cohen}, or uniaxial stress along the polar direction could help to bring
the free energies closer to each other.\cite{Budimir05}

\bigskip

Very recently, surprisingly high $d_{33}$ values of 2000 pC/N have been
obtained by Chu \textit{et al.} \cite{Chu} in BaTiO$_{3}$ poled along the
[720] direction. This direction is perpendicular to a cleavage plane and
allows high values of the electric field to be applied without fracturing
the crystal. This leads to a very large deformation and maximum strain
values of 0.6\% with fields below 10 kV/cm. However, with slight deviations
from this poling direction, the $d_{33}$ value is reduced by one-half.
Further theoretical and experimental work is needed to understand this
strikingly high values in the piezoelectric response of BaTiO$_{3}$.

\bigskip

From the above discussion, it is clear that, despite the many unknowns, our
understanding of the piezoelectric response in perovskites has improved
considerably in the last couple of years. Accordingly, although it is not
yet clear that a monoclinic unit cell is a necessary condition for large
piezoelectric responses, it seems nevertheless wise to focus the search for
highly responsive dielectric and piezoelectric systems around low symmetry
phases that will allow polarization rotation, namely, those polar phases
that, independent of the local symmetry, appear to be monoclinic or
triclinic from x-ray or neutron diffraction measurements. These phases will
be likely to appear in regions bridging higher symmetry phases of similar
energy.

\section{\protect\bigskip Acknowledgements}

It is with much sadness that we dedicate this article to the memory of Gen
Shirane. We had the great privilege of working closely with Gen on
structural investigations of the morphotropic phase boundaries in different
lead oxide systems during his last five years at Brookhaven National
Laboratory, and we have profited enormously from his deep insight into the
physics of the problem, and his unerring instinct for doing the right
experiment. Working with him was an entertaining, rewarding and
unforgettable experience!

We would like to thank Ilya Grinberg, Andrew Rappe and Yu Wang for sharing
with us their unpublished results, Ben Burton for an interesting
conversation about the importance of the order of the tetragonal-monoclinic
phase transition, and Mike Glazer for many illuminating discussions.
Financial support from the U. S. Department of Energy, Division of Materials
Sciences, under Contract No. DE-AC02-98CH10886, is also acknowledged.

\bigskip




\begin{thebibliography}{999}
\bibitem{Footnote} In conventional Rietveld analysis, the peak profile model
is assumed to have a smooth dependence on $2\theta $, whereas in practice it
is often found that there is additional broadening associated with different 
$hkl$ classes of reflections due to microstructural effects such as size,
strain and local order.
\end{thebibliography}
\end{document}